\documentclass[aps,prapplied,reprint,superscriptaddress,showpacs]{revtex4-2}
\usepackage{graphicx}
\usepackage{dcolumn}
\usepackage{bm}
 \usepackage{tensor}
 \usepackage{physics}
\usepackage{times}
\usepackage{siunitx}
\usepackage{xcolor}
\usepackage{setspace}

\begin{document}


\title{Multiprobe time reversal for high-fidelity vortex-mode-division multiplexing over a turbulent free-space link}

\author{Yiyu Zhou}
\email{yzhou62@ur.rochester.edu}
\affiliation{The Institute of Optics, University of Rochester, Rochester, New York 14627, USA}
\author{Jiapeng Zhao}
\affiliation{The Institute of Optics, University of Rochester, Rochester, New York 14627, USA}
\author{Boris Braverman}
\affiliation{Department of Physics, University of Ottawa, Ottawa, Ontario K1N 6N5, Canada}
\author{Kai Pang}
\affiliation{Department of Electrical and Computer Engineering, University of Southern California, Los Angeles, California 90089, USA}
\author{Runzhou~Zhang}
\affiliation{Department of Electrical and Computer Engineering, University of Southern California, Los Angeles, California 90089, USA}
\author{Alan E. Willner}
\affiliation{Department of Electrical and Computer Engineering, University of Southern California, Los Angeles, California 90089, USA}
\author{Zhimin Shi}
\affiliation{Department of Physics, University of South Florida, Tampa, Florida 33620, USA}
\author{Robert W. Boyd}
\affiliation{The Institute of Optics, University of Rochester, Rochester, New York 14627, USA}
\affiliation{Department of Physics, University of Ottawa, Ottawa, Ontario K1N 6N5, Canada}



\date{\today}

\begin{abstract}
The orbital angular momentum (OAM) of photons presents a degree of freedom for enhancing the secure key rate of free-space quantum key distribution (QKD) through mode-division multiplexing (MDM). However, atmospheric turbulence can lead to substantial modal crosstalk, which is a long-standing challenge to MDM for free-space QKD. Here, we show that the digital generation of time-reversed wavefronts for multiple probe beams is an effective method for mitigating atmospheric turbulence. We experimentally characterize seven OAM modes after propagation through a 340-m outdoor free-space link and observe an average modal crosstalk as low as 13.2\% by implementing real-time time reversal. The crosstalk can be further reduced to 3.4\% when adopting a mode spacing $\Delta \ell$ of 2. We implement a classical MDM system as a proof-of-principle demonstration, and the bit error rate is reduced from $3.6\times 10^{-3}$ to be less than $1.3\times 10^{-7}$ through the use of time reversal. We also propose a practical and scalable scheme for high-speed, mode-multiplexed QKD through a turbulent link. The modal crosstalk can be further reduced by using faster equipment. Our method can be useful to various free-space applications that require crosstalk suppression.
\end{abstract}
\maketitle

\section{Introduction}

\begin{figure}[b]
\center
\includegraphics[width= \linewidth]{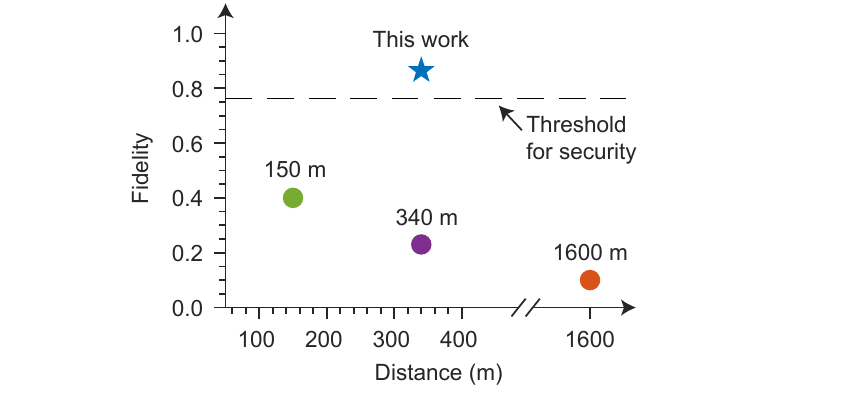}
\caption{The previously reported mode fidelity measured in 150- \cite{mphuthi2019free}{}, 340- \cite{zhao2020performance}{}, and 1600-m \cite{lavery2017free}{} links when using an OAM mode spacing $\Delta \ell=1$. The dashed line represents the fidelity threshold of 76.3\% for secure QKD with seven-dimensional encoding.}
\label{fig:compare}
\end{figure}

In recent decades, quantum key distribution (QKD) has attracted increasing interest because it can guarantee communication security based on fundamental laws of quantum mechanics\cite{gisin2002quantum}{}. Free-space QKD \cite{liao2017satellite,ursin2007entanglement, hughes2002practical, schmitt2007experimental, buttler1998practical}{} can guarantee communication security between mobile nodes such as aircraft and satellites. In addition, free space presents lower loss than fibers and thus is favorable to loss-sensitive applications such as quantum teleportation \cite{ren2017ground}{} and entanglement distribution \cite{yin2017satellite}{}. Because of multiple constraints imposed by quantum protocols (e.g., the use of low-brightness quantum light sources, high sensitivity to loss and crosstalk, etc.), the secure key rate of QKD is significantly lower than the data transfer rate of classical communication protocols. Thus, it remains highly desirable to enhance the secure key rate of QKD. The spatial degree of freedom is a promising candidate for boosting capacity of both quantum and classical communication through mode-division multiplexing (MDM) \cite{wang2012terabit, ren2016experimental, li2017high} or high-dimensional encoding \cite{krenn2014communication, trichili2016optical, mirhosseini2015high, sit2017high, mphuthi2019free, zhao2020performance, zhou2019using, krenn2016twisted} and is compatible with polarization- and wavelength-division multiplexing. In particular, slowly diverging spatial modes such as the vortex modes carrying orbital angular momentum (OAM) are commonly used as a basis set in free-space communication compared to alternative basis sets such as discrete spot arrays \cite{walborn2006quantum} that are unsuitable for a long-distance link. However, atmospheric turbulence inevitably leads to strong modal crosstalk between spatial modes \cite{paterson2005atmospheric, malik2012influence}{}, which severely degrades the channel capacity of a free-space link. We next summarize several previous works to show the typical level of crosstalk between OAM modes in a turbulent, outdoor free-space channel. In a 150-m link \cite{mphuthi2019free}{}, the crosstalk fluctuates between 60\% and 80\% depending on time. In a 340-m cross-campus link \cite{zhao2020performance}{}, the crosstalk is measured to be in the range between 70\% and 80\% when using a fast steering mirror and an adaptive optics system for compensation. In a 1600-m link \cite{lavery2017free}{}, the modal crosstalk can be as high as 90\%. Hence, the turbulence can be a serious concern for crosstalk-sensitive applications such as QKD. We summarize the results that have been previously reported with a mode spacing $\Delta \ell=1$ in Fig.~\ref{fig:compare}. A more comprehensive summary is listed in Table S1 within the Supplemental Material \cite{Supplement}{}.

\begin{figure*}[t]
\center
\includegraphics[width= \linewidth]{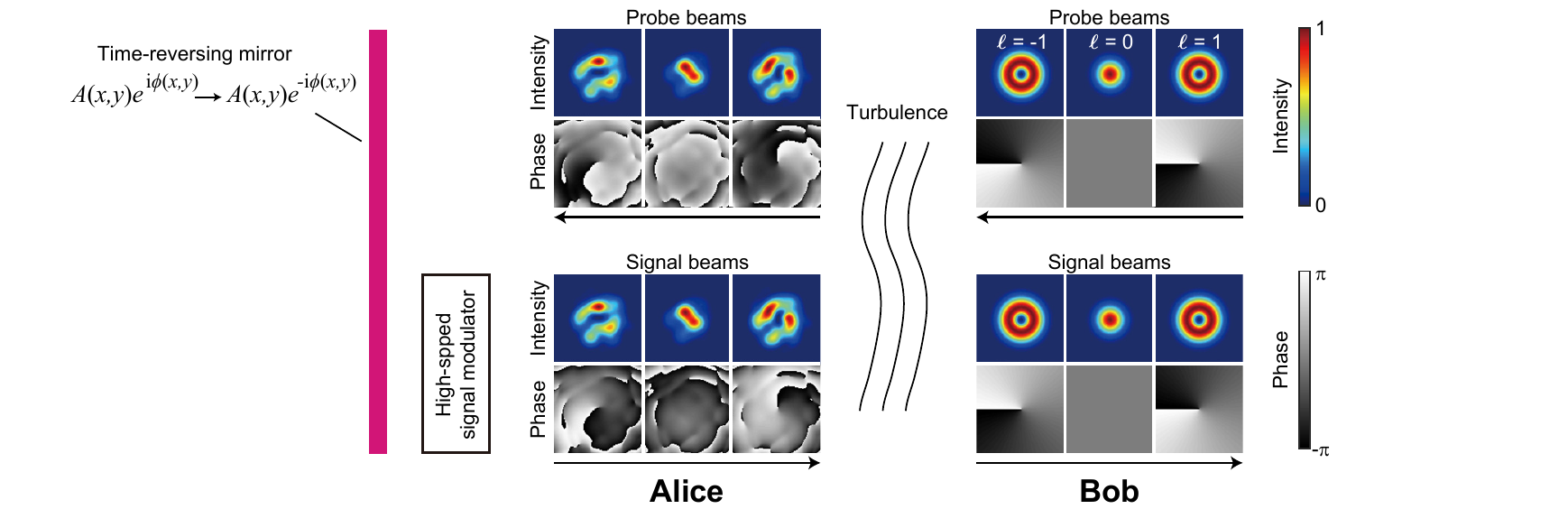}
\caption{Illustration of time reversal to transmit high-fidelity modes of $\ell=-1,0,1$. Alice uses a time-reversing mirror to generate the phase conjugates of the probe beams transmitted by Bob and sends it to Bob as signal beams. Information can be transferred from Alice to Bob by using a high-speed optical modulator to modulate the time-reversed beams at Alice's side. The arrows under the spatial modes indicate the propagation direction.}
\label{fig:turbulence}
\end{figure*}

Adaptive optics is the most common method for turbulence correction and has been widely adopted for astronomical imaging \cite{tyson2015principles}{}. A conventional adaptive optics system consists of a wavefront sensor and a deformable mirror at the receiver. The wavefront sensor measures the aberrated phase of an incoming beacon beam (typically a Gaussian beam), and subsequently the deformable mirror corrects the phase aberration of the incoming beam based on the feedback from wavefront sensor as post-turbulence compensation \cite{ren2014adaptiveOL}{}. However, different OAM modes exhibit mode-dependent amplitude and phase distortions after propagation through the same turbulent link due to the mode-dependent diffraction \cite{zhao2020performance}{}. Therefore, a conventional post-turbulence phase-only adaptive optics system is unable to correct the distortions for different OAM modes simultaneously, even in principle \cite{neo2014correcting, neo2016measurement}{}. In addition, the phase compensation imposed by the adaptive optics cannot correct the amplitude distortions, which inherently limits the performance of conventional adaptive optics. Furthermore, the effectiveness of adaptive optics for OAM communication has mostly been tested in numerical simulations \cite{neo2016measurement, li2014evaluation, zhao2012aberration}{} or in lab-scale links with emulated, slowly varying, fully controllable turbulence \cite{liu2019single, ren2014adaptive, li2016compensation, chen2016demonstration}{}. There is only one experimental demonstration using adaptive optics for OAM communication through an outdoor link \cite{zhao2020performance}{}, and the crosstalk is reduced from 80\% to 77\% by using both adaptive optics and a fast steering mirror simultaneously. This level of crosstalk is too large to guarantee secure QKD \cite{cerf2002security}{}. Therefore, adaptive optics has achieved very limited performance enhancement in outdoor free-space OAM communication links despite numerous simulations and lab-scale experiments. Other methods for modal crosstalk suppression, such as a multiple-input multiple-output (MIMO) algorithm \cite{ren2015free}{} and an artificial neural network (ANN) \cite{krenn2014communication}{}, cannot be applied to QKD because these algorithms require a large number of photons for digital signal processing and thus are inappropriate for quantum applications that operate at a single-photon level.

\begin{figure*}[t]
\center
\includegraphics[width= \linewidth]{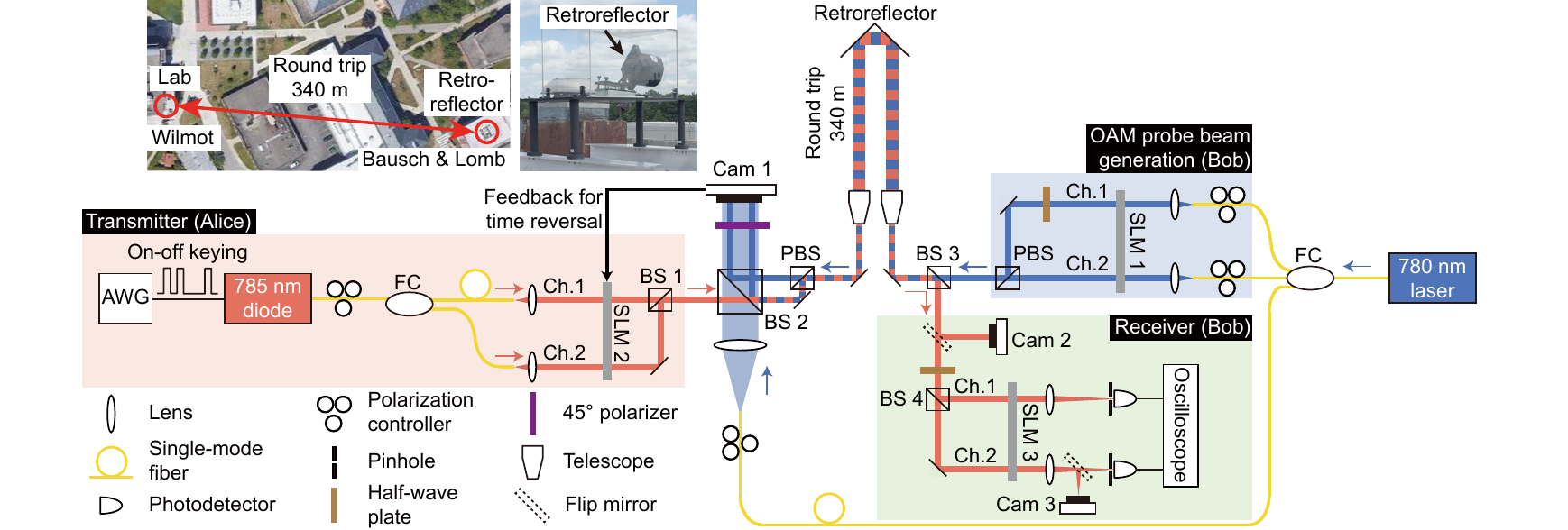}
\caption{Schematic of the experiment. The upper-left image shows an aerial view of the free-space link and is courtesy of Map data \textcopyright{}2020 Google. The retroreflector is installed on a building rooftop. Digital time reversal is realized by displaying a computer-generated diffractive hologram on SLM~2 based on the feedback from a camera (cam~1). AWG is the arbitrary waveform generator. BS is the beamsplitter. PBS is the polarizing beamsplitter. FC is the fiber coupler. Ch. 1 is channel 1. Ch. 2 is channel 2. Additional experimental details are presented in Sec. I of the Supplemental Material \cite{Supplement}{}.}
\label{fig:setup}
\end{figure*}

\section{Digital time reversal}
Time reversal is a well-established method for aberration cancelation and is also referred to as phase conjugation. Nonlinear optical effects such as four-wave mixing are the standard method to realize time reversal \cite{boyd2003nonlinear}{}. However, nonlinear time reversal can only transmit one spatial mode at a time and thus cannot be used to enable mode-division multiplexing. Time reversal has been applied in the time domain to cancel fiber nonlinearity \cite{liu2013phase}{}, in the polarization domain to cancel fiber birefringence \cite{muller1997plug}{}, and in the spatial domain for scattering suppression through biological tissues \cite{yaqoob2008optical, cui2010implementation}{} and multimode fibers \cite{papadopoulos2012focusing, bae2019compensation, zhou2020high, czarske2016transmission}{}. However, these demonstrations cannot be readily used to enable mode-division multiplexing over rapidly varying turbulence. As can be seen from two recent reviews on free-space optical communication \cite{cox2020structured, trichili2020roadmap}{}, adaptive optics remains the standard method for turbulence correction, while time reversal has not yet been considered for free-space communication. Here we propose and demonstrate that high-speed multiprobe time reversal combined with fast wavefront sensing can be used to effectively suppress atmospheric turbulence for OAM communication through a 340-m free-space link. Figure~\ref{fig:turbulence} illustrates how to transmit high-fidelity spatial modes from Alice to Bob by using time reversal with multiple probe beams. As shown in Fig.~\ref{fig:turbulence}, Bob first transmits standard OAM modes (Laguerre-Gauss modes with radial index $p=0$ and OAM index $\ell=-1,0,1$) to Alice, and the modes received by Alice are distorted both in amplitude and phase. By contrast, perfect OAM modes can be transmitted to Bob if Alice uses a time-reversing mirror. For an arbitrary incident spatial mode $A(x,y)e^{i\phi(x,y)}$, the mode reflected by a time-reversing mirror becomes $A(x,y)e^{-i\phi(x,y)}$. After propagation through the same turbulent link, the OAM modes received by Bob become the time reversal of the originally transmitted OAM modes and can in principle have a perfect spatial mode quality \cite{boyd2003nonlinear}{}, assuming a link without beam clipping. The time-reversing mirror can be digitally implemented by a phase-only spatial light modulator (SLM) \cite{cui2010implementation}{}, because spatial amplitude and phase profile modulation can be simultaneously realized by a diffractive hologram \cite{arrizon2007pixelated}{}. In order to compensate for time-varying turbulence, the hologram needs to be updated dynamically in real time. From a technical point of view, the spatial modes transmitted from Bob to Alice can be regarded as probe beams that enable fast characterization of turbulence and thus allow Alice to perform preturbulence mode generation for each spatial mode.

\section{Experiment}
Our time reversal experimental schematic is presented in Fig.~\ref{fig:setup}. We first characterize the modal crosstalk matrix of the time reversal system, and then use the setup to realize a two-channel OAM communication system. A static diffractive hologram is displayed on SLM~1 at Bob's side to generate two 780~nm OAM probe beams of horizontal and vertical polarization, respectively. These two OAM probe beams are combined by a PBS and then transmitted to Alice through a retroreflector. These two orthogonally polarized probe beams allow simultaneous digital time reversal for two different OAM modes, which facilitates the realization of two-channel OAM communication system discussed later. We refer to our method as multiprobe time reversal because multiple probe beams are needed to enable mode-division multiplexing. The retroreflector of 127~mm diameter is installed on a building rooftop that is 170~m away, resulting in a round trip distance of 340 m. The light beams do not hit the center of the retroreflector, and Alice's and Bob's telescope apertures are approximately 8 cm apart from each other. Therefore, the forward-propagating beam and the reflected, backward-propagating beam do not overlap with each other, and thus they experience different turbulence effects. At Alice's side, a PBS is used to separate the two aberrated OAM modes, and a coherent reference plane wave is combined with two separated OAM probe beams by a BS. A camera (cam~1) is used to record the interference fringes, and a 45$^{\circ}$ polarizer is inserted before cam~1 to enhance the interference pattern visibility. Through the standard off-axis holography analysis \cite{cuche2000spatial}{} (see Sec. I of the Supplemental Material \cite{Supplement}{}), the amplitude and phase profile of the received modes can be retrieved with a single-shot measurement from cam~1. Hence, cam~1 is used as a wavefront sensor in our setup. To provide the coherent reference plane wave for off-axis holography, a single-mode fiber is used to guide the continuous-wave 780~nm light source from Bob to Alice. We emphasize that this single-mode fiber can be avoided by using alternative wavefront sensors such as a Shack-Hartmann sensor \cite{abado2010two}{} or complex field direct measurement \cite{shi2015scan}{}. Based on the measured amplitude and phase profile of the aberrated OAM modes, Alice computes the corresponding diffractive hologram and displays it on SLM~2 to generate the time reversals of the received OAM modes with desired amplitude and phase profile at the first diffraction order (see Sec. I of the Supplemental Material \cite{Supplement}{}). Alice uses a 785~nm laser diode as the light source and transmits the time-reversed modes to Bob. Because of the negligible dispersion of free space, the difference in wavefront distortion between 780 and 785~nm wavelength can be ignored \cite{zhao2020performance}{}. Furthermore, the use of different wavelengths makes it possible to use dichroic mirrors to separate forward-propagating and backward-propagating beams with little loss, which is crucial to loss-sensitive applications such as QKD. In the experiment we use seven Laguerre-Gauss modes of $\ell=-3,-2,\cdots, 3$ to test the performance of the time reversal system. The aperture diameter of both telescopes is 5~cm, resulting in a Fresnel number product of $N_f=D^2/\lambda z=9.4$, where $D=5$~cm, $\lambda=780$~nm, and $z=340$~m. The beam waist radius of the OAM modes is $w_0=10$~mm after beam expansion of the telescope. The turbulence structure constant \cite{andrews2005laser}{} $C_n^2$ is measured to be in the range of $2.2 \times 10^{-15}$ to $8.6\times 10^{-15}$~m$^{-2/3}$, the Fried parameter $r_0$ ranges from 0.16 to 0.07~m, and thus $D/r_0$ is between 0.31 and 0.70. Additional experimental details can be found in Sec. I of the Supplemental Material \cite{Supplement}{}.

\begin{figure*}[t]
\center
\includegraphics[width= \linewidth]{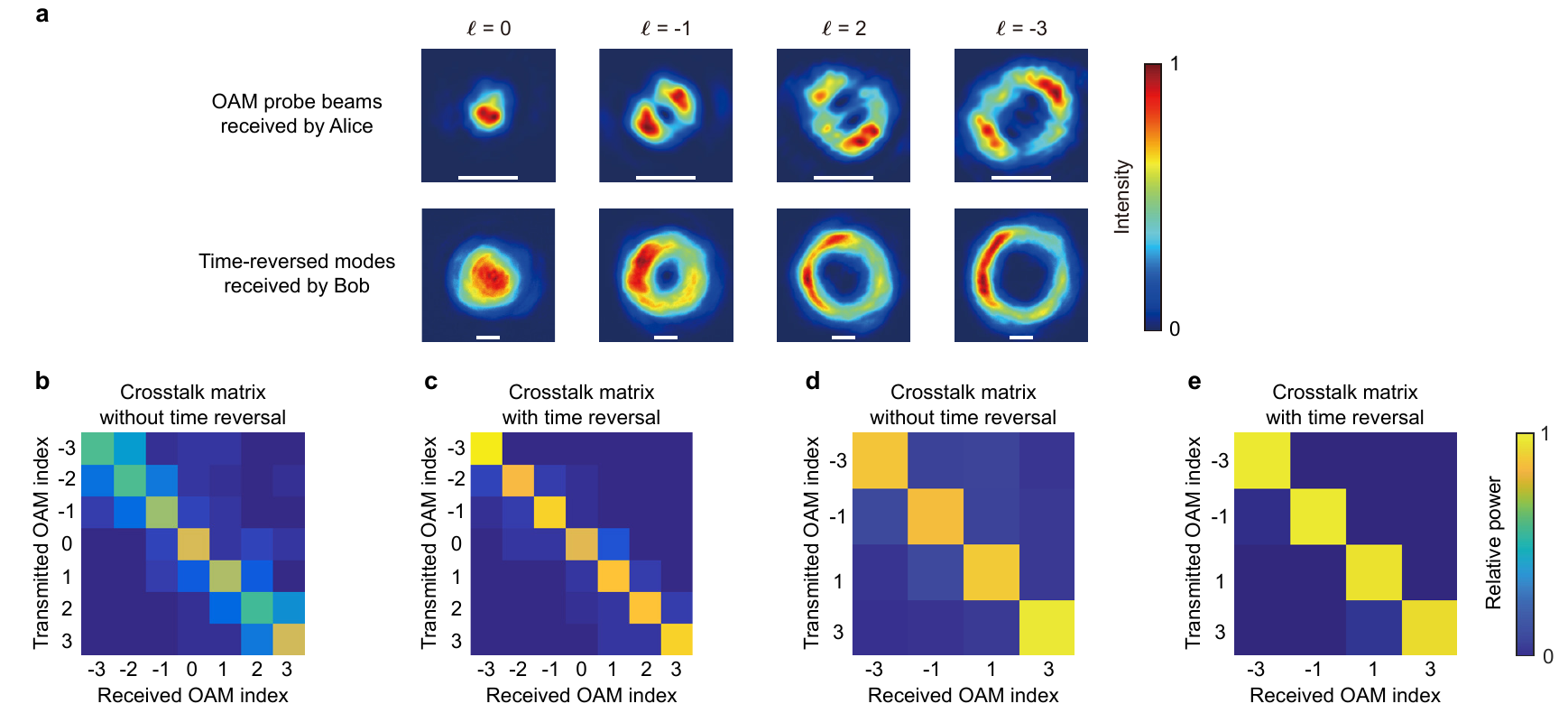}
\caption{(a) Typical examples of the spatial modes received by Alice (top row) and Bob (bottom row). Scale bar is 0.5~mm. (b) Experimentally measured crosstalk matrix without time reversal. The average crosstalk is 37.0\%. (c) Crosstalk matrix with time reversal. The average crosstalk is 13.2\%. (d) Crosstalk matrix using a mode spacing $\Delta \ell$ of 2 without time reversal. The average crosstalk becomes 10.0\%. (e) Crosstalk matrix using a mode spacing $\Delta \ell$ of 2 with time reversal. The average crosstalk is reduced to 3.4\%.}
\label{fig:Image}
\end{figure*}

\section{Results}

Because of atmospheric turbulence and aberration of the telescope system, the OAM probe beams received by cam~1 at Alice's side exhibit clear distortions as shown in the top row of Fig.~\ref{fig:Image}(a). Alice generates the time reversals of these aberrated modes and transmits them to Bob. The time-reversed modes received by cam~2 at Bob's side are shown in the bottom row of Fig.~\ref{fig:Image}(a), which exhibit improved mode fidelity compared to those received by Alice. To quantify the modal crosstalk of the time-reversed modes, we display a densely encoded diffractive hologram on SLM~3 as a mode demultiplexer \cite{trichili2016optical}{}, and a camera (cam~3) is placed at the Fourier plane of SLM~3 to measure the crosstalk matrix \cite{mphuthi2019free}{} (see Sec. I of the Supplemental Material \cite{Supplement}{}). When Alice transmits standard OAM modes, the crosstalk matrix of the modes received by Bob is shown in Fig.~\ref{fig:Image}(b). The average fidelity (i.e. the value of the diagonal elements) is 63.0\%, and thus the average crosstalk is 37.0\%. Therefore, this link cannot support secure QKD with OAM encoding in the absence of time reversal since the crosstalk is higher than the security error threshold of 23.7\% for a seven-dimensional system \cite{cerf2002security}{}. By contrast, when Alice transmits time-reversed OAM modes to Bob, the average crosstalk is reduced to 13.2\% and thus the fidelity becomes 86.8\% as shown in Fig.~\ref{fig:Image}(c), which allows for secure QKD operating at a single-photon level. In addition, it is well known that the modal crosstalk can be reduced by increasing the mode spacing $\Delta \ell$ \cite{malik2012influence}{}. We calculate the crosstalk matrix with a mode spacing $\Delta \ell$ of 2 by postselecting the data of $\ell=-3,-1,1,3$, and the average crosstalk can be further reduced from 10.0\% to 3.4\% by using time reversal as shown in Fig.~\ref{fig:Image}(d,e). These results are the lowest crosstalk ever achieved in an outdoor free-space link (see Table S1 within the Supplemental Material \cite{Supplement}{} for a comprehensive comparison to other previous works). Although we use classical light to characterize the crosstalk matrix, we emphasize that our method can be readily applied to quantum applications that operate at a single-photon level.

Ideally, a realization of time reversal can completely eliminate the aberration and achieve zero crosstalk. Here we attribute the nonzero crosstalk observed in our experiment to the following reasons. First, the operational bandwidth of our digital time reversal system is limited. The image transfer time from cam~1 to computer memory is 4~ms, the computation time for diffractive hologram generation is 1~ms, and the refresh time of SLM~2 is 5~ms, resulting in a total response time of approximately 10~ms and hence an operational bandwidth of 100~Hz. By contrast, the characteristic frequency of turbulence can be tens of to hundreds of hertz \cite{tyson2015principles}{}. We believe that the operational bandwidth of digital time reversal can be improved to exceed 1~kHz with faster devices such as a high-bandwidth Shack-Hartmann wavefront sensor \cite{abado2010two}{} and a 22~kHz SLM \cite{ren2015tailoring}{}. Potential advantages and disadvantages of using faster devices are discussed in Supplemental Material Section I \cite{Supplement}. Second, the mode fidelity of our time reversal generation is not perfect. The time reversal generation fidelity can be further improved by calibrating and correcting the residual aberration of the SLM. Third, beam clipping should be avoided in a free-space link in order to enable a perfect realization of time reversal. In our experiment, beam clipping can occur at the telescope aperture as well as at the retroreflector, and the link transmittance for $\ell=\pm 3$ is approximately 10\% lower than that of $\ell=0$ (see Sec. II of the Supplemental Material \cite{Supplement}{}). It should be noted that the retroreflector is unnecessary and can be removed in a realistic free-space link. In addition, low-cost and large-diameter Fresnel lenses (1~m diameter lens is commercially available \cite{yang2019realization}{}) can be used in a long-distance link to avoid beam clipping. The static aberration of the Fresnel lens should not be a concern because it can be corrected by time reversal as part of the overall channel aberration.
\begin{figure}[tb]
\center
\includegraphics[width=\linewidth]{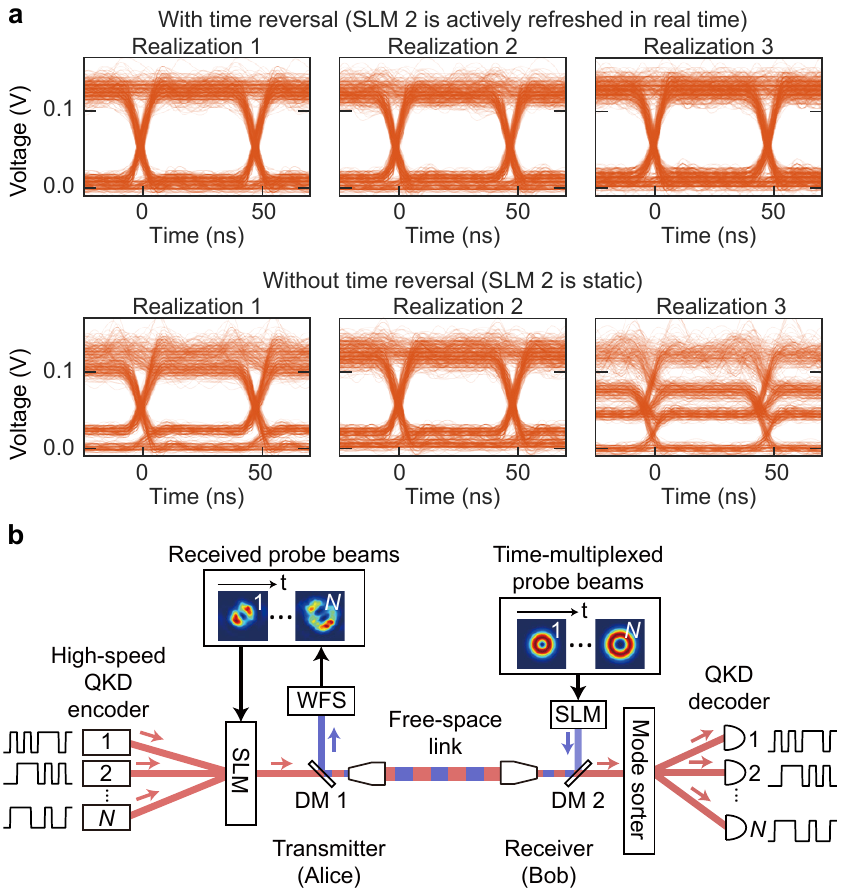}
\caption{(a) Eye diagrams of different realizations when transmitting the $\ell=2$ and $\ell=3$ modes and receiving the $\ell=2$ mode. The diagrams with time reversal are displayed in the top row, and those without time reversal are shown in the bottom row. (b) Proposed free-space QKD system with $N$-channel OAM multiplexing using time reversal for modal crosstalk suppression. The quantum signal streams are represented by binary on-off keying signals for simplicity. DM is the dichroic mirror. WFS is the wavefront sensor.}
\label{fig:classical}
\end{figure}

It should be noted that the data transfer rate of OAM communication is not limited by the SLM refresh rate but is decided by the modulation speed of the transmitter \cite{chen2016demonstration}{}. To clarify this point, we demonstrate a proof-of-principle classical communication system with two-channel OAM multiplexing. We choose $\ell=2$ for channel 1 and $\ell=3$ for channel 2. Bob generates the horizontally polarized $\ell=3$ mode and vertically polarized $\ell=2$ mode and transmits them to Alice. It should be noted that the polarization has negligible effect on beam propagation due to the small birefringence of turbulent free space. Hence, Alice can measure two aberrated OAM modes simultaneously with a single-shot measurement. We emphasize that even though we are using orthogonally polarized OAM modes to facilitate turbulence characterization, this is unnecessary and can be avoided by transmitting different OAM modes in different time slots to Alice using a spatial mode switch \cite{braverman2020fast}{}. The intensity of the 785~nm laser diode at Alice's side is modulated at the rate of 20~Mbit/s with an on-off keying format. The modulation rate is solely limited by our modulator bandwidth and can be readily improved to the Gbit/s level by using commercially available high-speed modulators. A fiber coupler is used to split the beam, and a 10-m fiber delay line is used to decorrelate the signal streams. The two beams illuminate separate areas of SLM~2 to generate the corresponding time reversal of the two OAM modes transmitted by Bob. The two time-reversed OAM modes are combined by BS~1 and then transmitted to Bob. It should be noted that both time-reversed modes are vertically polarized, and the horizontal polarization is an unused degree of freedom that can be further adopted for polarization encoding or multiplexing if needed. In the schematic shown in Fig.~\ref{fig:setup}, we use BS~4 at the receiver to split the beam and subsequently use SLM~3 to perform projection onto two OAM modes. In the experiment we omit BS~4 and perform different OAM projections by switching the hologram for simplicity. The projection measurement realized by SLM~3 can be replaced by a low-loss OAM mode sorter \cite{mirhosseini2013efficient}{} for loss-sensitive applications such as QKD. Within every 4~s we collect $8\times 10^4$ bits from the oscilloscope, and we collect a total of $8\times 10^6$ bits over 400~s. The eye diagrams at different times for $\ell=2$ are shown in Fig.~\ref{fig:classical}(a). Alice first implements time reversal by dynamically refreshing SLM~2 in real time, and the corresponding eye diagrams are shown in the top row. As a comparison, Alice also imprints a static hologram on SLM~2 to transmit standard OAM modes to Bob without implementing time reversal, and the corresponding eye diagrams are shown in the bottom row. In both cases, the bit rate of received signals is 20 Mbit/s. It can be seen that the bit rate is determined by the intensity modulator at Alice's side and is not limited by the refresh rate of the SLM. A slight power fluctuation can be observed in the eye diagrams between different realizations, and we attribute it to the laser diode power instability as well as the SLM phase flickering. In addition, we can see that the signal streams with time reversal exhibit lower crosstalk than those without time reversal. Off-line digital signal processing is performed to analyze the signal, and the average bit error rate is $3.6\times 10^{-3}$ without time reversal. By contrast, we detected 0 bit errors when time reversal is performed, implying a bit error rate lower than $1.3\times 10^{-7}$. It should be noted that the bit error rate discussed in this work is different from the quantum bit error rate that is defined for the detection of a single photon. In a classical communication system, the bit error rate is dependent on channel crosstalk and the signal-to-noise ratio \cite{sarkar2009study}{}. The minimum measurable bit error rate is restricted by the memory capacity of the oscilloscope in our experiment. Time-resolved bit-error-rate measurements for both channels are presented in Sec. III of the Supplemental Material \cite{Supplement}{}. We believe that gigahertz and higher modulation speeds can be achieved simply by using faster modulators and detectors \cite{wang2012terabit}{}. Finally, we wish to emphasize that the previously demonstrated lab-scale adaptive optics system with preturbulence compensation by Ren \textit{et al.} \cite{ren2014adaptive}{} is fundamentally different from our scheme for two reasons. First, we transmit different OAM modes as probe beams, allowing for mode-dependent pre-turbulence compensations for individual modes. By contrast, Ren \textit{et al.} always used a Gaussian mode with $\ell=0$ as the probe beam. Second, we use diffractive holograms on a SLM to control both the amplitude and phase profile of each time-reversed mode, while Ren \textit{et al.} use a SLM to apply a phase-only compensation to all different OAM modes simultaneously. Because of these limitations, the preturbulence compensation demonstrated in Ref. \cite{ren2014adaptive}{} is essentially equivalent to the conventional post-turbulence compensation with back-propagating beams and thus does not exhibit a better performance than the conventional adaptive optics.

Based on the low crosstalk and high communication bandwidth of the system, here we propose a practical, scalable free-space QKD system with $N$-channel OAM multiplexing using time reversal for modal crosstalk suppression. The schematic of the proposed QKD system is shown in Fig.~\ref{fig:classical}(b), and the procedure to implement time reversal can be summarized by the following steps. (i) At Bob's side, a SLM and a laser light source are used to generate and switch among $N$ standard OAM modes sequentially. (ii) Alice uses a wavefront sensor to measure the amplitude and phase profile of each OAM mode in real time. (iii) A densely encoded diffractive hologram \cite{trichili2016optical}{} is computed and imprinted onto a SLM at Alice's side to generate and multiplex $N$ time-reversed modes simultaneously. The times needed to implement steps (i), (ii), and (iii) are denoted by $t_1$, $t_2$, and $t_3$, respectively. The operational bandwidth of the digital time reversal system, defined as $1/(t_1+t_2+t_3)$, needs to be higher than the turbulence characteristic frequency as discussed earlier. It can be seen that $t_1$ and $t_2$ are proportional to $N$, and thus high-speed, low-latency devices are needed if $N$ is large. We note that high-speed OAM mode-switching to accelerate step (i) can be readily achieved by using a digital micromirror device at 22~kHz \cite{ren2015tailoring}{} or by an acousto-optic modulator at 500~kHz \cite{braverman2020fast}{}. A 100~kHz wavefront sensor has also been reported \cite{abado2010two}{}, which can be used to accelerate step (ii). High-performance digital circuits such as field-programmable gate arrays and digital signal processors can be used to accelerate step (iii). An alternative method to significantly reduce $t_1$ and $t_2$ is to generate OAM modes of different wavelengths at Bob's side. These OAM modes can be separated from each other at Alice's side by using gratings or dichroic mirrors, and thus the need of high-speed mode switching can be eliminated. Although the densely encoded hologram typically has a low diffraction efficiency, this is not a problem for coherent-state-based QKD protocols, because strong loss is inherently needed to attenuate a classical, high-brightness laser to a single-photon level. The standard polarization-encoded decoy-state QKD protocol \cite{schmitt2007experimental}{} can be used to enable secure communications. In this case, information can be encoded onto photons by using a high-speed polarization modulator at Alice's side, and the secure key rate of each channel is not limited by the SLM refresh rate but determined by the polarization modulation rate that can readily reach the gigahertz level \cite{li2019high}{}. In fact, by adding a high-speed polarization switch and attenuating the laser diode to a single-photon level, our classical MDM system can be immediately turned to a polarization-encoded OAM-multiplexed QKD system. Alternative coherent-state-based protocols such as time-bin encoding \cite{islam2017provably}{} and continuous-variable encoding \cite{grosshans2003quantum}{} are also applicable to our scheme. Furthermore, wavelength-division multiplexing is compatible with our scheme because of the nondispersive, broadband spectral response of free space. The major advantage of this time reversal QKD protocol is the low crosstalk as demonstrated in our experiment, which has not been achieved by any adaptive optics system in an outdoor turbulent link and there is still room for improvement. Moreover, this protocol cannot be replaced by classical modal crosstalk suppression methods such as a MIMO and an ANN for quantum applications operating at a single-photon level as discussed earlier.

\section{Conclusion}
In conclusion, we experimentally demonstrate effective modal crosstalk suppression in a 340-m free-space OAM communication link through the use of digital time reversal. The crosstalk induced by turbulence can be reduced from 37.0\% to 13.2\% with time reversal, and further down to 3.4\% by using a mode spacing $\Delta \ell$ of 2. We believe that lower crosstalk can be reasonably achieved by using faster equipment in a straightforward manner. A proof-of-principle classical communication system is realized to show the feasibility of high-speed communication with OAM multiplexing. In addition, a practical and scalable scheme for free-space QKD with OAM multiplexing is also proposed and analyzed. Based upon the scalability of the experimental implementation and low crosstalk of the data, we anticipate that digital time reversal can be useful to numerous free-space quantum and classical applications that require modal crosstalk suppression.

\section*{Acknowledgements}
This work is supported by the U.S. Office of Naval Research (Grants No. N00014-17-1-2443, No. N00014-20-1-2558, and No. N00014-16-1-2813). B.B. acknowledges the support of the Banting Postdoctoral Fellowship. R.W.B. acknowledges funding from the Natural Sciences and Engineering Research Council of Canada, the Canada Research Chairs program, and the Canada First Research Excellence Fund award on Transformative Quantum Technologies. A.E.W. acknowledges the support of the Vannevar  Bush  Faculty  Fellowship  sponsored  by  the  Basic Research Office of the Assistant Secretary of Defense for Research and Engineering. R.Z. acknowledges the support of the Qualcomm Innovation Fellowship. Y.Z. acknowledges the technical support from Wenxiang Hu, Myron W. Culver, and Christopher Harvey.

%

\clearpage
\widetext
\begin{center}
\Large{\textbf{Supplemental Material}}
\end{center}

\setstretch{1.25}
\setcounter{equation}{0} \setcounter{subsection}{0} \setcounter{section}{0}
\setcounter{figure}{0}
\renewcommand{\theequation}{S\arabic{equation}}
\renewcommand{\thefigure}{S\arabic{figure}}
\renewcommand{\thetable}{S\arabic{table}}
\renewcommand{\refname}{Supplemental References}
\renewcommand{\bibnumfmt}[1]{[S#1]} 
\renewcommand{\citenumfont}[1]{S#1}

\section{Experimental setup}

\begin{figure}[h]
\center
\includegraphics[width= 1\linewidth]{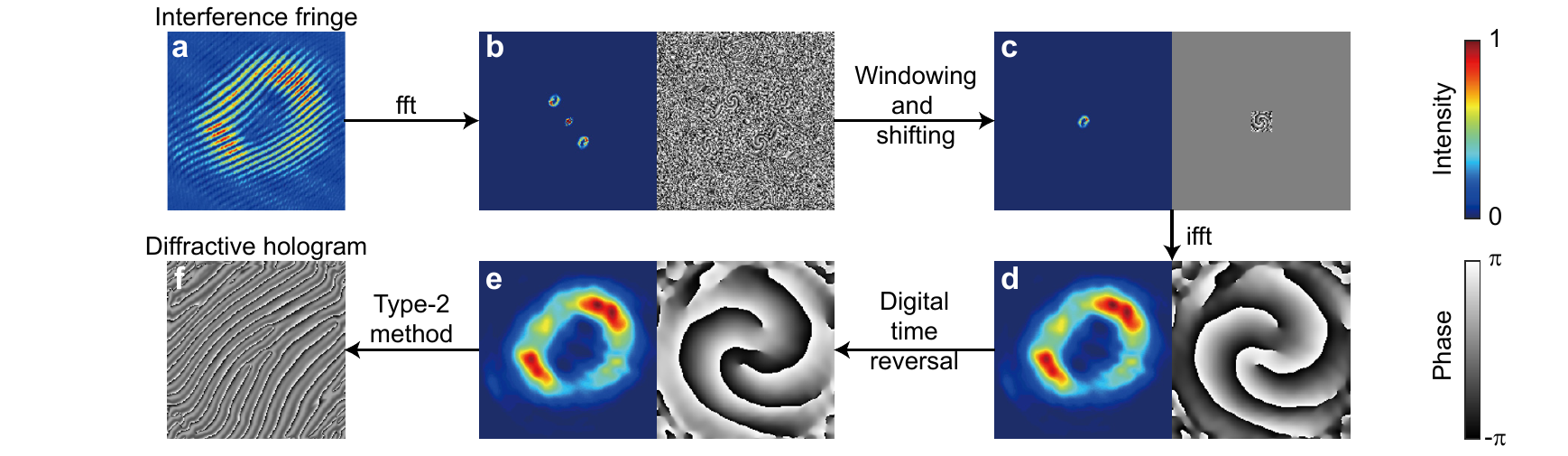}
\caption{Procedure for digital time reversal. (a) The interference fringe pattern recorded by Cam~1. (b) The fast Fourier transform (fft) of interference fringe. (c) The first order is selected by a window and shifted to the center. (d) The inverse fast Fourier transform (ifft) gives the amplitude and phase of the received probe beam. (e) The digital time reversal is implemented by flipping the sign of the phase. (f) The computer-generated phase-only diffractive hologram that can generate the time-reversed OAM mode at the first diffraction order.}
\label{fig:SuppHOLO}
\end{figure}

\begin{figure}[h]
\center
\includegraphics[width= 1\linewidth]{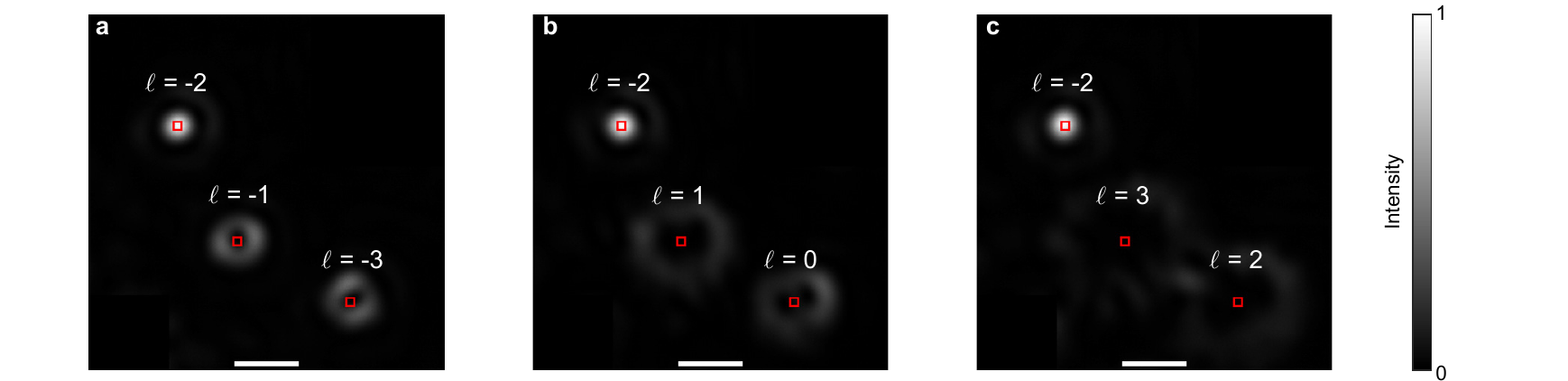}
\caption{Experimentally recorded images for crosstalk matrix measurement using mode-multiplexed computer-generated holograms. The input beam is a time-reversed $\ell=-2$ mode. The hologram performs OAM projection measurements onto (a) $\ell=-2, -1, -3$, (b) $\ell=-2,1,0$, and (c) $\ell=-2,3,2$, respectively. The red boxes denote the windows used for crosstalk matrix measurement. Scale bar: 0.5~mm.}
\label{fig:SuppDEMUX}
\end{figure}

In the experiment, Bob uses a 780~nm laser (DL pro, Toptica) as light source and a phase-only SLM (Pluto 2, Holoeye) as SLM~1. The beam waist radius of the OAM modes before Bob's telescope is $\omega_0=0.7$~mm. The telescope at Bob's side consists of a $f=35$~mm lens (LA1027-B, Thorlabs) and a $f=500$~mm lens (LA1380-AB, Thorlabs). The retroreflector (\#49-672, Edmund Optics) is installed on a building roof, and the total round trip distance is 340~m. The telescope at Alice's side consists of a $f=11$~mm lens (C220TMD-B, Thorlabs) and a $f=400$~mm lens (LA1725-A, Thorlabs). Cam~1 (BFS-U3-04S2M-CS, FLIR) is used to acquire the image of received OAM modes. Alice uses Cuda C++ language on a desktop computer (CPU: Intel i7-9700K, GPU: Nvidia RTX 2070 Super) to perform off-axis holography analysis, compute the hologram (type-2 hologram in ref. \cite{Sarrizon2007pixelated}{}), and display the hologram on SLM~2 (HSP-1920-488-800, Meadowlark Optics). The computational procedure for off-axis holography and digital time reversal is presented in Fig.~\ref{fig:SuppHOLO}. One example of the interference fringes recorded by Alice is shown in Fig.~\ref{fig:SuppHOLO}(a). We perform fast Fourier transform of the interference fringe, select the first order and shift it to the image center. By performing an inverse fast Fourier transform, the amplitude and phase of the received probe beam can be obtained. Digital time reversal is implemented by flipping the sign of the phase. Then we use the type-2 method in \cite{Sarrizon2007pixelated}{} to generate the phase-only diffractive hologram, which can be used to generate the time reversal of the probe beam at the first diffraction order with desired amplitude and phase. A 785~nm laser diode (LP785-SF20, Thorlabs) is used as Alice's light source. The time-reversed OAM modes generated by SLM~2 are transmitted to Bob through the same free-space link, and Bob uses Cam~2 (BFS-U3-16S2M-CS, FLIR) to record images of the received modes. Bob uses a separate active area on SLM~2 as the independent SLM~3 to implement spatial mode demultiplexing. Mode-multiplexed holograms \cite{Smphuthi2019free, Strichili2016optical}{} are used to project an optical beam onto three different OAM modes simultaneously. We use Cam~3 (BFS-U3-16S2M-CS, FLIR) to measure the diffraction orders at the Fourier plane of SLM~3, and a few experimentally recorded images are shown in Fig.~\ref{fig:SuppDEMUX} for a time-reversed $\ell=-2$ mode. We show the 1st-order diffractions and other irrelevant diffraction orders are blocked. For each diffraction order, we use a window of $7\times 7$ pixel size at the beam center as depicted by the red boxes in Fig.~\ref{fig:SuppDEMUX} and calculate the total power inside the window. We capture 500 images for each hologram to get the statistical average of the relative power ratio between different modes. For each transmitted mode, three mode-multiplexed holograms are used to measure the crosstalk spectrum for $\ell=-3,-2,\cdots, 3$. We repeat this procedure for seven OAM modes to get a $7\times 7$ crosstalk matrix. The optical signals with on-off-keying modulation are converted to voltage waveforms by using a photodetector (APD130A, Thorlabs) connected to an oscilloscope. For each received voltage waveform, we specify a voltage threshold and convert the signals to binary bit streams by comparing the signals to the threshold. The voltage threshold is tuned independently for each signal stream to reach the minimized the bit error rate.

In our experiment, the bit rate is limited by the bandwidth of intensity modulator, and the mode fidelity is partly limited by the operational bandwidth of the digital time reversal system. We use 20 Mbps on-off keying intensity modulation to realize a classical mode-division multiplexing system. Commercially available, state-of-the-art modulators can readily achieve a rate well above Gbps. The main advantage of using high-speed modulators is that the bit rate can be boosted by orders of magnitude. However, the electronic driver may not maintain signal integrity in high-speed applications. If signal waveforms are distorted severely at a high frequency, the consequent disadvantage is that the bit error rate will increase. We use a camera as the wavefront sensor, and the time needed to transfer one frame of image from the camera to the computer in our setup is approximately 4~ms. If a higher-speed wavefront sensor is available, the operational bandwidth of the entire time reversal system can be increased. The consequent advantage is that our system can compensate for faster turbulence in a longer link and also can allow for multiplexing more OAM modes. On the other hand, a high-speed wavefront sensor usually has a lower signal to noise ratio and a smaller pixel count. Therefore, the precision of wavefront sensing can be reduced. As a result, the mode fidelity of time-reversed modes might decrease. Furthermore, we use a phase-only liquid-crystal-on-silicon SLM with a refresh rate of approximately 200~Hz to digitally generate time-reversed modes. To achieve a higher refresh rate, one can use a digital micromirror device (DMD). The advantage of a DMD is that it can have a significantly higher refresh rate as large as 22 kHz \cite{Sren2015tailoring}{}. The disadvantage of using DMD is that its diffraction efficiency is low. The low diffraction efficiency of DMD is not a problem to coherent-state-based QKD protocols because strong loss is inherently needed to attenuate the laser light to a single-photon level. However, for classical communications, the low efficiency results in a low signal to noise ratio and thus might increase bit error rate.

\clearpage
\section{Transmittance of free-space link}
\begin{figure}[ht]
\center
\includegraphics[width= 1\linewidth]{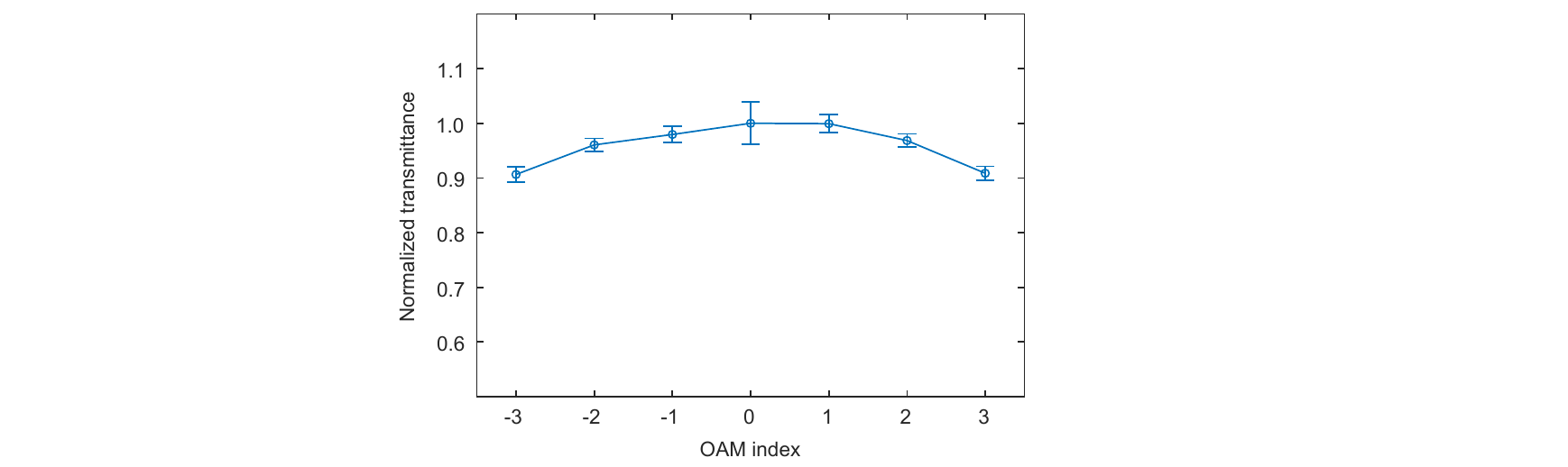}
\caption{The transmittance of the free-space link for different OAM modes. The average transmittance for $\ell=0$ is normalized to unity. The error bar represents one standard deviation.}
\label{fig:Transmittance}
\end{figure}

\clearpage
\section{Bit error rate measurement}

\begin{figure}[ht]
\center
\includegraphics[width= 1\linewidth]{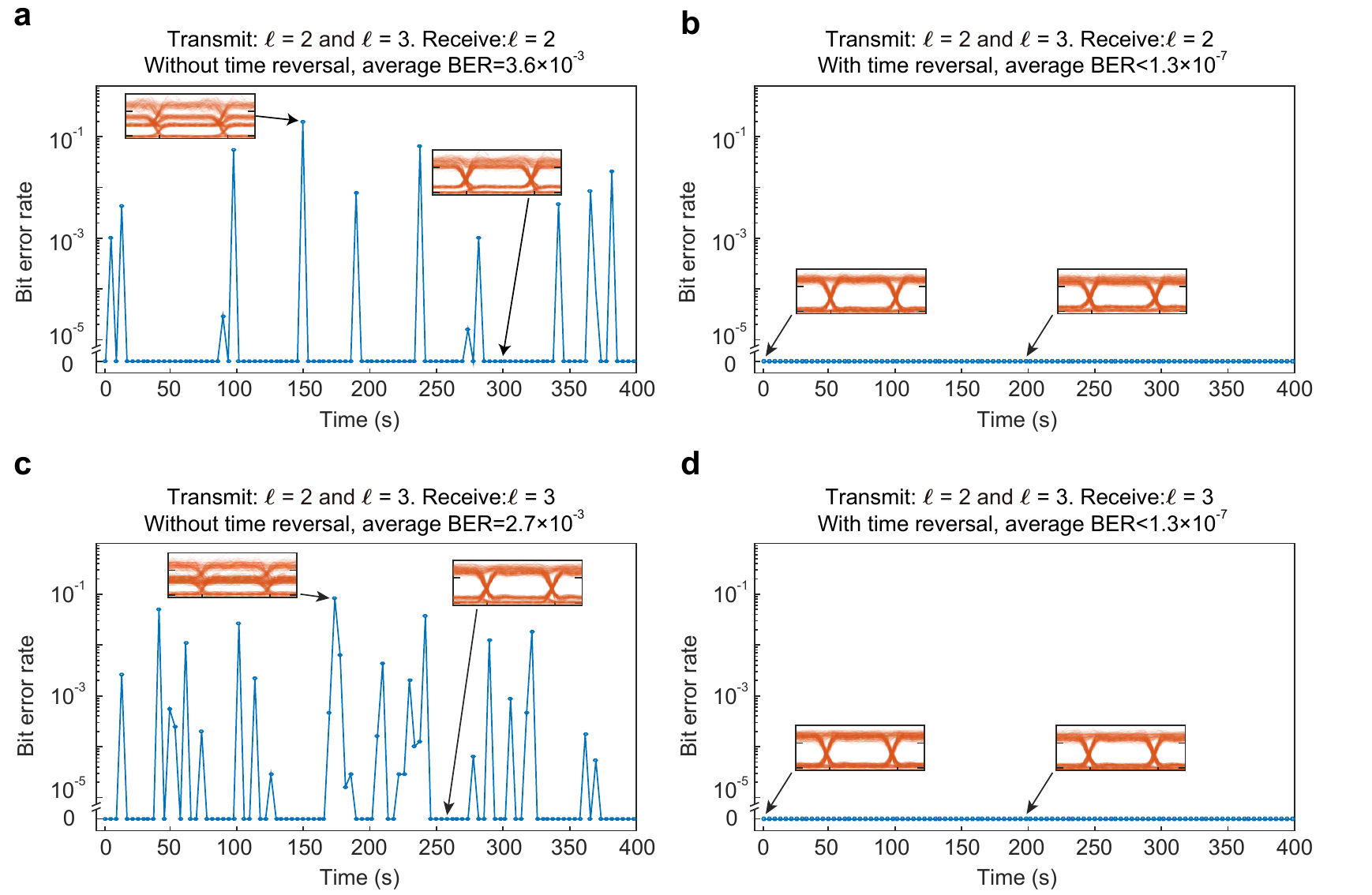}
\caption{Time-resolved bit error rate (BER) measurement. The transmitted OAM modes are $\ell=2$ and $\ell=3$. (a) The time-resolved BER for receiving $\ell=2$ mode without time reversal. The average BER is $3.6\times 10^{-3}$. (b) The time-resolved BER for receiving $\ell=2$ mode with time reversal. We detected 0 bit error within $8\times 10^6$ received bits with time reversal, and therefore the average BER is less than $1.3\times 10^{-7}$. (c) The time-resolved BER for receiving $\ell=3$ mode without time reversal. The average BER is $2.7\times 10^{-3}$ without time reversal. (d) The time-resolved BER for receiving $\ell=3$ mode with time reversal. We detected 0 bit error within $8\times 10^6$ received bits with time reversal, and therefore the average BER is less than $1.3\times 10^{-7}$. The inset shows the eye diagrams acquired by oscilloscope.}
\label{fig:BER1}
\end{figure}

\clearpage
\section{Summary of previous works}

\begin{table}[h]

\centering
\begin{tabular}{|c|c|c|c|c|}
\hline
No. & Link length                                                 & Parameters                                                               & Results                                                                                                                                                        & Compensation method                                                                \\ \hline
1   & 100 m \cite{Sli2017high}                                                       & \begin{tabular}[c]{@{}c@{}}$\lambda$=1550 nm,\\ large $N_f$\end{tabular} & \begin{tabular}[c]{@{}c@{}}Crosstalk$\approx$20\% with\\ a mode spacing $\Delta \ell$ of 2\end{tabular}                                                                   & Fast steering mirror                                                               \\ \hline
2   & 120 m \cite{Sren2016experimental}                                                      & \begin{tabular}[c]{@{}c@{}}$\lambda$=1550 nm\\ $N_f$=20.8\end{tabular}   & \begin{tabular}[c]{@{}c@{}}Crosstalk=4.8\% with\\ a mode spacing $\Delta \ell$ of 2\end{tabular}                                                                               & None                                                                               \\ \hline
3   & 150 m \cite{Smphuthi2019free}                                                      & \begin{tabular}[c]{@{}c@{}}$\lambda$=532 nm\\ $N_f$=97.0\end{tabular}    & \begin{tabular}[c]{@{}c@{}}Crosstalk$\approx$60\%\\ during the daytime and\\ $\approx$80\% during the nighttime\\ with a mode spacing $\Delta \ell$ of 1\end{tabular}          & None                                                                               \\ \hline
4   & 300 m \cite{Ssit2017high}                                                      & \begin{tabular}[c]{@{}c@{}}$\lambda$=850 nm\\ $N_f$=22.1\end{tabular}    & \begin{tabular}[c]{@{}c@{}}Crosstalk=11\% with\\ a mode spacing $\Delta \ell$ of 4\end{tabular}                                                                                & \begin{tabular}[c]{@{}c@{}}Post-selecting lucky\\ beams\end{tabular}               \\ \hline
5   & 340 m \cite{Szhao2020performance}                                                      & \begin{tabular}[c]{@{}c@{}}$\lambda$=633 nm\\ $N_f$=4.89\end{tabular}    & \begin{tabular}[c]{@{}c@{}}Crosstalk can be reduced\\ from 80\% to 77\%\\ by using compensation\\ with a mode spacing $\Delta \ell$ of 1\end{tabular}                          & \begin{tabular}[c]{@{}c@{}}Adaptive optics\\ and fast steering mirror\end{tabular} \\ \hline
6   & 1.6 km \cite{Slavery2017free}                                                     & \begin{tabular}[c]{@{}c@{}}$\lambda$=809 nm\\ $N_f$=8.7\end{tabular}     & \begin{tabular}[c]{@{}c@{}}Crosstalk$\approx$90\%\\ with a mode spacing $\Delta \ell$ of 1\end{tabular}                                                                        & None                                                                               \\ \hline
7   & 3 km \cite{Skrenn2014communication}                                                      & $\lambda$=532 nm                                                         & Bit error rate=$1.7\times 10^{-2}$                                                                                                                                          & Artificial neural network                                                                                \\ \hline
8   & 143 km \cite{Skrenn2016twisted}                                                     & $\lambda$=532 nm                                                         & Bit error rate=$8.3\times 10^{-2}$                                                                                                                                          & Artificial neural network                                                                           \\ \hline
9  & \begin{tabular}[c]{@{}c@{}}340 m\\ (this work)\end{tabular} & \begin{tabular}[c]{@{}c@{}}$\lambda$=780 nm\\ $N_f$=9.4\end{tabular}     & \begin{tabular}[c]{@{}c@{}}Crosstalk reduced from\\ 37.0\% to 13.2\% with \\ a mode spacing $\Delta \ell$ of 1, and\\ from 10.0\% to 3.4\% with\\ a mode spacing $\Delta \ell$ of 2\end{tabular} & Digital time reversal
                                                                 \\ \hline
\end{tabular}
\caption{A summary of previous works for free-space OAM communications. $\lambda$: wavelength. $N_f$: Fresnel number product. It can be seen that only our time reversal method presents significant modal crosstalk suppression, and our experiment has the lowest modal crosstalk. It should be noted that the works in No. 7 and No. 8 report their results in terms of bit error rate rather than crosstalk. In addition, the artificial neural network used in these works requires data acquisition using a camera. Therefore, the communication speed is fundamentally limited by the camera image acquisition rate, which is on the order of kHz. Furthermore, since a high signal-to-noise ratio is needed for image processing, this method cannot be directly used in quantum communications operating at a single-photon level.}
\end{table}

\clearpage

\end{document}